\documentclass[10pt,conference,onecolumn]{IEEEtran}
\usepackage[utf8]{inputenc}
\usepackage{paralist}
\usepackage{algpseudocode}
\usepackage{graphicx}
\usepackage{color}
\usepackage{amsfonts}
\usepackage{amssymb}
\usepackage{mathtools}
\usepackage{xspace}

\newtheorem{theorem}{Theorem}

\newcommand{\eo}{\textsf{EOp}\xspace}

\IEEEoverridecommandlockouts
\title{Mining Network Events using Traceroute Empathy%
\thanks{Work partly supported by the European Community's Seventh 
Framework Programme (FP7/2007-2013), grant no. 317647 (``Leone'' project).}}

\author{\IEEEauthorblockN{Marco {Di Bartolomeo}\IEEEauthorrefmark{1}, Valentino {Di Donato}\IEEEauthorrefmark{1}, Maurizio
 Pizzonia\IEEEauthorrefmark{1}, Claudio Squarcella\IEEEauthorrefmark{1}, Massimo Rimondini\IEEEauthorrefmark{1}}
 \IEEEauthorblockA{\IEEEauthorrefmark{1}Roma Tre University, Department of 
Engineering
    \\\{dibartolomeo, didonato, pizzonia, squarcel, rimondin\}@dia.uniroma3.it}
}

\graphicspath{{figures/}}

\mathchardef\t="2D

\mathchardef\breakingcomma\mathcode`\,
{\catcode`,=\active
  \gdef,{\breakingcomma\discretionary{}{}{}}
}

\let\circwithoutspace\circ
\renewcommand{\circ}{\,\circwithoutspace\,}

\newcommand{\deltapre}[2]{\delta_{#1}^\mathrm{pre}(#2)}
\newcommand{\deltapost}[2]{\delta_{#1}^\mathrm{post}(#2)}

\newcommand{\preemp}[1]{\stackrel{\mathclap{\mathrm{pre}}}{\sim}_{#1}}
\newcommand{\postemp}[1]{\stackrel{\mathclap{\mathrm{post}}}{\sim}_{#1}}

\newcommand{\updown}{{\downarrow\!\uparrow}}

\newcommand{\pre}{\textsf{pre}}
\newcommand{\post}{\textsf{post}}

\begin{document}
\maketitle

\begin{abstract}

With the increasing diffusion of Internet probing technologies,
a large amount of regularly collected traceroutes are available for
Internet Service Providers (ISPs) at low cost. We introduce a practically 
applicable methodology and algorithm that, given 
solely an arbitrary set of traceroutes, spot routing paths that change 
similarly over time, aggregate them into inferred events, and report each event 
along with the impacted observation points and a small set of IP addresses that 
can help identify its cause.
The formal model at the basis of our methodology revolves around the notion of 
\emph{empathy}, a relation that binds similarly behaving traceroutes. 
The correctness and completeness of our approach are based on structural 
properties that are easily expressed in terms of empathic measurements.
We perform experiments with data from public measurement infrastructures like 
RIPE Atlas, showing the effectiveness of our algorithm in distilling 
events from a large amount of traceroute data. We also validate the accuracy of 
the inferred events against ground-truth knowledge of routing changes 
originating 
from induced and spontaneous routing events.
Given these promising results, we believe our methodology can be an effective aid 
for ISPs to detect and track routing changes affecting many users (with 
potentially adverse effects on their connection quality).

\end{abstract}
\section{Introduction}

One of the primary goals of a network operator is to ensure its network works as expected.
Since misbehaviors can happen for a variety of reasons, constant monitoring is
performed by operators to timely detect problems and limit users complaints.
Directly monitoring the health of each network element works in many situations, 
but may fall short when the element itself lacks the necessary agent support or 
is not under the operator's control. Also, there are cases in which network 
elements are reported as working despite end-to-end communication being impaired 
by misconfigurations or subtle hardware failures (the
\emph{silent failures} in~\cite{kompella2007detection}).

A large corpus of research works has focused on methodologies to detect and 
locate faults using information collected by hardware or software elements
(called \emph{monitors} or \emph{probes}) deployed throughout 
the network, possibly far away from the problem.
Indeed, large probing networks are already running to ease the assessment of 
service levels by regulators (e.g.,~\cite{samknows})
or for management and scientific purposes (e.g.,~\cite{atlas,ark,mlab}). 
A widely discussed approach to localize the faults on a network consists in 
correlating end-to-end measurements from a large number of probes using a 
technique called \emph{binary tomography}~\cite{duffield2006network, 
netDiagnoser,kompella2007detection}. However, several problems hinder the 
application of this approach in a production 
environment~\cite{cunha2009measurement,huang2008practical}: false positives in 
the detection phase, the failure to accommodate network dynamics, and
the need for a complete knowledge of the topology and for a synchronization among 
the probes.
Other approaches take advantage of control plane 
information~\cite{jcckak-pircipc-2013,fmzbm-liri-2004}, which may require a 
complex collecting infrastructure or may just not be available to the operator 
for the part of the network that is not under his control.

In this paper we introduce a novel methodology and an algorithm that enable the 
analysis of large collections of traceroute measurements in search of significant 
changes, thus easing management and troubleshooting. 
Our methodology takes as input only a set of traceroutes, identifies 
paths that evolve similarly over time, and reports them aggregated into 
inferred 
events (e.g., routing changes, loss of connectivity), augmented with an impact 
estimate and a restricted set of IP addresses that are likely close to the cause 
of the event (a piece of information similar to those provided by tomography-based 
techniques). The methodology, as well as its correctness, are founded on the notion of 
\emph{empathy}, a relation that binds similarly behaving traceroutes, which are 
therefore a good evidence of the same network event.
Our approach does not need a-priori knowledge of the network topology, 
does not 
assume a stable routing state, and does not impose restrictions on the schedule 
of traceroutes, which may be collected asynchronously and at arbitrary 
intervals. Instead, it takes advantage of asynchronous 
measurements to improve 
the timeliness and precision of event detection, and is almost unaffected by 
measurement errors (e.g., due to software errors or routing anomalies), which in 
most cases only generate fictitious events with a small impact.
 
We provide experimental evidence of the effectiveness of our 
approach by running our algorithm on data collected by large-scale measurement 
infrastructures such as RIPE Atlas, and by comparing the inferred events with 
ground truth derived from induced routing changes or third-party information.

The rest of the paper is organized as follows. In Section~\ref{sec:related-work} 
we review related contributions. In Section~\ref{sec:model} we describe
our network model and the fundamental properties of empathy. In
Section~\ref{sec:methodology} we introduce a methodology and an 
algorithm, based on empathy, to infer events and report relevant data about them.
In Section~\ref{sec:experiments} we analyze the results of the application of our 
methodology to real-world data. In 
Section~\ref{sec:futurework} we draw conclusions and present ideas for future 
work.

\section{Related Work}\label{sec:related-work}

There are many large-scale platforms that collect traceroute measurements 
(e.g.,~\cite{ark,atlas,samknows}), and a standardization effort is also 
ongoing~\cite{bagnulo2013standardizing}. As a consequence, there is a growing 
interest in finding patterns in such measurements, as confirmed, for 
example, in~\cite{brownlee2014searching}: in this paper the authors 
search for events by clustering data from~\cite{atlas} according to distance 
metrics that determine the amount of difference between subsequent 
traceroutes. While we also aim at grouping traceroute changes into 
events, our approach is based on the novel concept of empathy and is 
provably correct and complete.

A large number of contributions focus on identifying the location or the root
cause of a fault based on data gathered by measurement networks.
The binary tomography approach, firstly proposed for 
trees~\cite{duffield2006network}
and then extended to general 
topologies~\cite{netDiagnoser,kompella2007detection},
has applicability problems which have been discussed 
in~\cite{cunha2009measurement,huang2008practical,ma2014node}. Most notably, the 
authors assume at least a partial knowledge of the network topology (which must 
often be 
inferred from input data).
Similar approaches to root cause analysis have also been described for 
interdomain routing data~\cite{fmzbm-liri-2004,jcckak-pircipc-2013}.
A number of systems combine control plane information with data plane 
measurements: for example, Hubble~\cite{katz2008studying}, 
LIFEGUARD~\cite{katz2012lifeguard}, 
NetDiagnoser~\cite{netDiagnoser}, as well as~\cite{kompella2007detection}.
Our approach relies on traceroutes only, does not assume any knowledge of the 
network topology, and does not impose restrictions on the schedule of 
traceroutes.

\section{The Empathy Relationship}\label{sec:model}
In this section we describe the model we use to analyze traceroute 
paths and we introduce the concept of \emph{empathy}, which is at the basis of 
our event inference method. 

Let $G=(V,E)$ be a graph that models an IP network: vertices in $V$ are network 
devices (routers or end systems), and edges in $E$ are links between devices.
Some devices in $V$, called \emph{network probes} or \emph{sources}, periodically 
perform traceroutes towards a predefined set of \emph{destinations}. We assume 
that each traceroute is acyclic (otherwise there is evidence of a network 
anomaly) and instantaneous (reasonable because in the vast majority of cases 
traceroutes terminate within a smaller time scale than that of routing changes).

Let $i=(s,d)$, where $s\in V$ is a source and $d\in V$ is a 
destination.
A \emph{traceroute path} $p_i(t)$ measured at time $t$ by $s$ towards $d$ is a 
sequence
$\left<v_1\ v_2\ \dots\ v_n\right>$
such that $v_1=s$, $v_j\in V$ for $j=1,\dots,n$, and there is an edge in $E$ for 
each pair $(v_k,v_{k+1})$,
While we include the source in $p_i(t)$,
the destination may not appear because a traceroute may end at an unintended 
vertex different from $d$.
For convenience, let $V(p)$ be the set 
of vertices of path $p$.

\begin{figure}
   \centering
   \includegraphics[scale=.85]{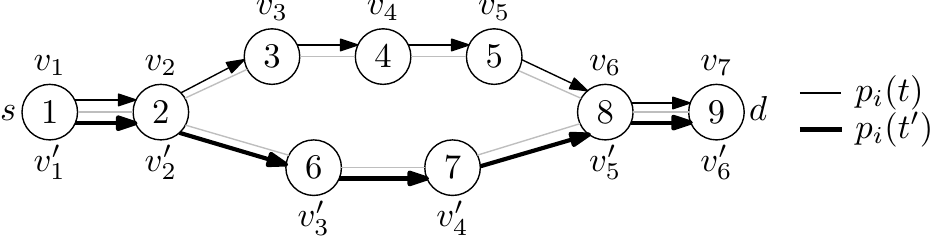}
   \caption{An example of two traceroute paths from $s$ to $d$ collected at 
different time instants $t$ and $t'$. Gray 
   lines represent network links.}
   \label{fig:path-example}
\end{figure}

Now, consider two traceroute paths
$p_i(t)=\left<v_1\ v_2\ \dots\ v_n\right>$ and
$p_i(t')=\left<v'_1\ v'_2\ \dots\ v'_m\right>$ between the same 
source-destination pair $i=(s,d)$, with $t'>t$,
and assume that $p_i(t)\neq p_i(t')$. Fig.~\ref{fig:path-example} shows two 
such 
paths: $i=(1,9)$,
$p_i(t)=\left<1\ 2\ 3\ 4\ 5\ 8\ 9\right>$, and
$p_i(t')=\left<1\ 2\ 6\ 7\ 8\ 9\right>$. Since the path from $s$ to $d$ has 
changed between $t$ 
and $t'$, we call the pair consisting of 
$p_i(t)$ and $p_i(t')$ a \emph{transition}, indicated by $\tau_i$, and say that 
it is \emph{active} at any time between the \emph{endpoints} $t$ and 
$t'$, excluding $t'$.
To analyze the path change, we focus on the portion of the two paths 
that has changed in the transition:
let $\deltapre{}{\tau_i}$ indicate the shortest subpath of $p_i(t)$ that goes 
from a vertex $u$ to a vertex $v$ such that all the vertices between 
$s$ and $u$ and between $v$ and the end of the path are unchanged in 
$p_i(t')$. If there is no such $v$ (for example because the destination is 
unreachable at $t$ or $t'$), $\deltapre{}{\tau_i}$ goes from $u$ to the end of 
$p_i(t)$.
Referring to the example in Fig.~\ref{fig:path-example}, it is
$\deltapre{}{\tau_i}=\left<2\ 3\ 4\ 5\ 8\right>$.
We define $\deltapost{}{\tau_i}$ as an analogous subpath of $p_i(t')$. Referring 
again to Fig.~\ref{fig:path-example}, it is
$\deltapost{}{\tau_i}=\left<2\ 6\ 7\ 8\right>$. In principle, 
$\deltapre{}{\tau_i}$ may have several vertices in common 
with $\deltapost{}{\tau_i}$ besides the first and the last one: we still 
consider $\deltapre{}{\tau_i}$ as a single continuous subpath, with negligible 
impact on the effectiveness of our methodology. The same applies 
to $\deltapost{}{\tau_i}$.
We can now introduce the concept of empathy, that determines when two traceroute 
paths exhibit a similar behavior over time.
Consider two transitions $\tau_1$, with source $s_1$ and destination $d_1$, and 
$\tau_2$, with source $s_2$ and destination $d_2$, such that both 
transitions are active between $t$ and $t'$, $t'>t$, at least one has an 
endpoint in 
$t$, and at least one has an endpoint in $t'$.
We say that \emph{$(s_1,d_1)$ and $(s_2,d_2)$ are pre-empathic at any time 
$t\leq \hat t<t'$}
if the portions of $p_1(t)$ and $p_2(t)$ that change during $\tau_1$ and 
$\tau_2$ 
overlap, namely $V(\deltapre{}{\tau_1})\cap V(\deltapre{}{\tau_2})\neq\emptyset$.
Intuitively, traceroute paths relative to pre-empathic sd-pairs stop 
traversing a network portion that they shared before an event occurred.
Similarly, we say that \emph{$(s_1,d_1)$ and $(s_2,d_2)$ are 
post-empathic at any time $t\leq\hat t<t'$}
if $V(\deltapost{}{\tau_1})\cap 
V(\deltapost{}{\tau_2})\neq\emptyset$.
Post-empathy captures a different kind of path change: traceroute paths 
of post-empathic sd-pairs start traversing a common portion that they did not 
use before the event occurred.
\begin{figure}
   \centering
   \includegraphics[scale=.8]{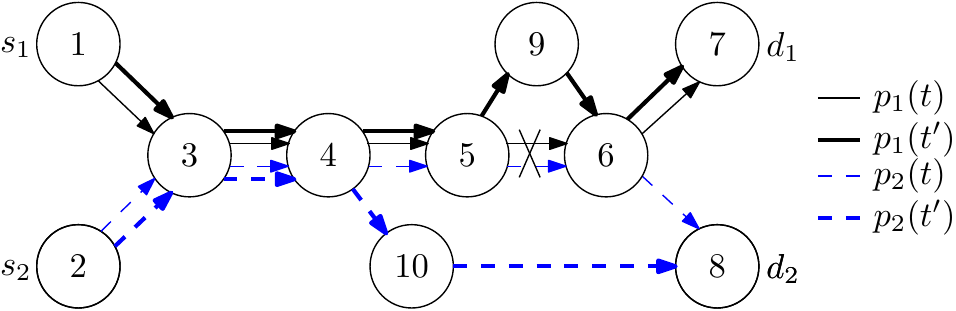}
   \caption{An example showing empathy relations. In this 
scenario link $(5,6)$ fails, and we have $(s_1,d_1)\preemp{t}(s_2,d_2)$ but 
$(s_1,d_1)\not\postemp{t}(s_2,d_2)$.}
   \label{fig:empathy_examples}
\end{figure}
Fig.~\ref{fig:empathy_examples} shows two traceroute paths 
$p_1$, from $s_1$ to $d_1$, and $p_2$, from $s_2$ to $d_2$, that change between 
$t$ and $t'$ due to the failure of link $(5,6)$. Considering the corresponding 
transitions $\tau_1$ and $\tau_2$, we have 
$\deltapre{}{\tau_1}=\left<5\ 6\right>$,
$\deltapost{}{\tau_1}=\left<5\ 9\ 6\right>$,
$\deltapre{}{\tau_2}=\left<4\ 5\ 6\ 8\right>$, and
$\deltapost{}{\tau_2}=\left<4\ 10\ 8\right>$. Since $\deltapre{}{\tau_1}$ and 
$\deltapre{}{\tau_2}$ share vertices $5$ and $6$, $(s_1,d_1)$ and $(s_2,d_2)$ 
are pre-empathic between $t$ and $t'$, 
whereas 
$(s_1,d_1)$ and 
$(s_2,d_2)$ are not post-empathic despite the fact that $p_1(t')$ and $p_2(t')$ 
share a subpath.
Indeed, $p_1$ and $p_2$ behave similarly before 
the link fails and change to two independent routes afterwards.

In order to understand how empathy is useful to infer network events, we need to 
formally introduce the notion of event, qualifying it as physical to 
distinguish it from events inferred by our algorithm.
We call \emph{physical event} at time $\bar t$ the simultaneous disappearance of 
a set $E^\downarrow$ of links from $E$ (\emph{down event}) or the simultaneous 
appearance of a set $E^\uparrow$ of links in $E$ (\emph{up event}), such that:
\begin{itemize}
\item either $E^\downarrow=\emptyset$ or $E^\uparrow=\emptyset$ (a physical event 
is either the disappearance or the appearance of links, not both);
\item $E^\downarrow\subseteq E$ (only existing links can disappear);
\item $E^\uparrow\cap E=\emptyset$ (only new links can appear);
\item $\exists v\in V\mid\forall (u,w)\in E^\downarrow: u=v\textrm{ or }w=v$, 
and the same holds for $E^\uparrow$ (all disappeared/appeared edges have 
one endpoint vertex in common). Vertex $v$ is called \emph{hub} of the 
event (an event involving a single edge $(u,v)$ has two hubs: $u$ and $v$; any 
other event has a unique hub).
\end{itemize}
When the type of an event is not relevant, we 
indicate it as $E^{\updown}$.
This event model captures the circumstance in which one or more links attached 
to a network device fail or are brought up, including the case in which a whole 
device fails or is activated. Such events may 
be caused, for example, by failures of network interface cards, line cards, or 
routers, by accidental link cuts, by provisioning processes, and 
by administrative reconfigurations. Congestion is normally not among 
these causes, but may be detected as an event if it makes a balancer shift 
traffic away from a set of links.
Failures or activations of links that do not 
have a vertex in common are considered distinct events.
\normalmarginpar
We only consider \emph{visible} physical events, namely those that cause at least 
a transition to occur. Moreover,
we assume that every transition comprises at least one edge involved in an 
event, an assumption that is long-argued in the literature about root-cause 
analysis (see, e.g.,~\cite{jcckak-pircipc-2013}) and yet we deem reasonable 
because our goal is to detect events, not reconstruct their original cause.
Given a physical event $E^\updown$ occurred at time $\bar t$,
we define the \emph{scope} $S(E^\updown)$ of $E^\updown$ as the set of sd-pairs 
$i=(s,d)$ involved in the transitions that are active at $\bar t$. We also call 
\emph{impact} of $E^\updown$ the cardinality of $S(E^\updown)$.

Intuitively, our algorithm infers network events based on 
observed transitions and on empathy relationships that bind the involved 
sd-pairs: empathies are considered a good evidence of path changes that are due 
to the same physical event.
\section{Seeking Events: Methodology and Algorithm}\label{sec:methodology}

\renewcommand{\algorithmicrequire}{\textbf{Input:}}
\renewcommand{\algorithmicensure}{\textbf{Output:}}

\makeatletter
\newlength{\trianglerightwidth}
\settowidth{\trianglerightwidth}{$\triangleright$~}
\algnewcommand{\LineComment}[1]{\Statex \hskip\ALG@thistlm $\triangleright$ #1}
\algnewcommand{\LineCommentCont}[1]{\Statex \hskip\ALG@thistlm%
  \parbox[t]{\dimexpr\linewidth-\ALG@thistlm}{\hangindent=\trianglerightwidth \hangafter=1 \strut$\triangleright$ #1\strut}}
\makeatother

\algrenewcommand\Return{\State \algorithmicreturn{} }%

In this section, we describe our inference algorithm for detecting 
routing events.
The algorithm takes as input a set of traceroute paths, and produces as result 
a list of inferred events, each equipped with the following information:
\begin{inparaenum}[i)]
\item an interval of time in which the event is supposed to have occurred,
\item a set of sd-pairs affected by the event (the scope),
\item the type (\textsf{up}, \textsf{down}) of the event (when inferred), and
\item a set of IP addresses that, after the event has occurred, (dis)appeared in 
all the traceroutes performed between sd-pairs in its scope (these IPs are good 
hints for identifying the cause of the event).
\end{inparaenum}

We refer to the model illustrated in the previous section, considering the 
general case of non-synchronized traceroute measurements. That is, for an sd-pair 
$i=(s,d)$ traceroute paths $p_i(t)$ are only available at specific 
time instants $t\in\mathbb{R}$ that depend on $s$ (if probes are 
synchronized via NTP, whose precision is high enough for our needs, we can refer 
to a universal clock).
As we will show later, unsynchronized traceroutes can improve the 
accuracy of the interval reported by our algorithm for an inferred event.
For convenience, for a transition $\tau_i$ we define the \emph{changed set} 
$\Delta(\tau_i)$ consisting of \emph{extended addresses}, namely IP 
addresses in $V(\deltapre{}{\tau_i})$ labeled with a tag \pre{} and IP addresses 
in $V(\deltapost{}{\tau_i})$ 
labeled with a tag \post{}.

Our algorithm consists of three phases.

\begin{figure}
   \centering
   \includegraphics[width=0.6\columnwidth]{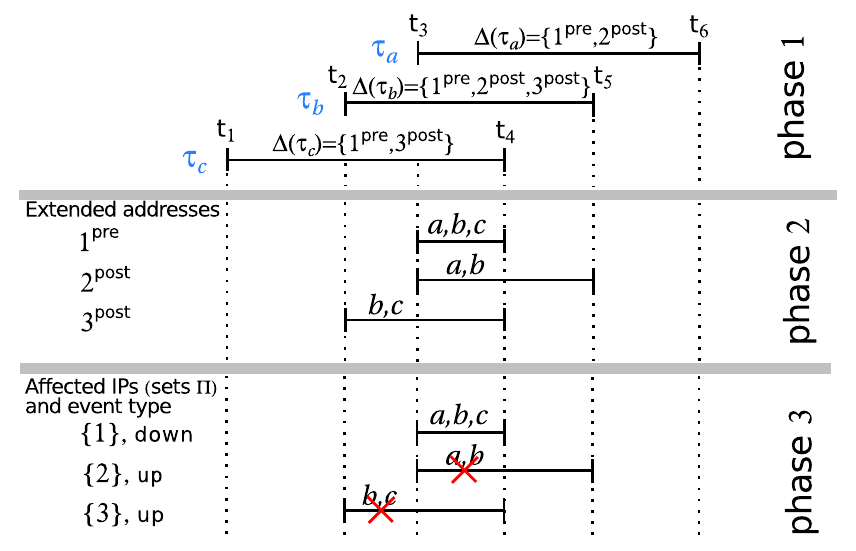}
   \caption{Sample outputs of the various phases of our algorithm.}
   \label{fig:algorithm_example}
\end{figure}

\emph{Phase 1 -- Identification of transitions}: in this phase, for each 
sd-pair $i$, input samples $p_i(t)$ are scanned and all 
transitions $\tau_i$, with the corresponding $\Delta(\tau_i)$,
are identified. The upper part of Fig.~\ref{fig:algorithm_example} shows an 
example with 3 transitions $\tau_a$, $\tau_b$, and $\tau_c$, represented 
as segments terminating at the transitions' endpoints, and 
the corresponding changed sets (IP addresses are represented as numbers).
The transitions in the figure can be the consequence of a physical down event 
with hub $1$.


\begin{figure}
\begin{algorithmic}[1]
\footnotesize
\Require a set $T$ of transitions
\Ensure a set $\mathit{CEvents}$ of candidate events, namely tuples $(t_1, t_2, 
S, \cal A)$ 
indicating time intervals in which all the sd-pairs in $S$ are pre-empathic 
or post-empathic with each other and all the corresponding transitions $\tau_i$ 
have 
extended address $\cal A$ in their changed set $\Delta(\tau_i)$.

\LineCommentCont{$S_{\cal A}$ and $t_{\cal A}$ are special variables for which 
the last two assigned values can be accessed by $prev()$ and $pprev()$ (if 
unset, they are $\emptyset$ and $-\infty$).}

\State $E$ = $\emptyset$
\For{every endpoint $t$ of transitions in $T$, in order of time}
\label{alg:endpoints}
   \For{every address $\cal A$ in the changed set of any transitions}
   \label{alg:extended-address}
      \State $T_{\cal A}$ = set of transitions $\tau_i$ active at $t$ such 
that ${\cal A}\in\Delta(\tau_i)$\label{alg:empathic-transitions}
      \State $sdp$ = the union of sd-pairs of transitions in 
$T_{\cal A}$
      \label{alg:empathy-sdpairs}
      \State \textbf{if} $S_{\cal A} \neq sdp$ \textbf{then} $S_{\cal A} = 
sdp$\label{alg:empathic-pairs}; $t_{\cal A}$ = $t$
      \If{$|pprev(S_{\cal A})| \leq |prev(S_{\cal A})|$ and $|prev(S_{\cal 
A})|>|S_{\cal A}|$}
      \label{alg:local-maximum}
         \State add $(prev(t_{\cal A}), t_{\cal A}, prev(S_{\cal A}), {\cal 
A})$ to 
$\mathit{CEvents}$
         \label{alg:add-candidate-event}
      \EndIf
   \EndFor
\EndFor
\Return $\mathit{CEvents}$
\end{algorithmic}
\caption{Phase 2 of our algorithm: it computes candidate events by finding sets 
of sd-pairs that are all pre-empathic or post-empathic with each other.}
\label{fi:algo:phase2}
\end{figure}

\emph{Phase 2 -- Construction of candidate events}: in 
this phase, the algorithm tracks the evolution of empathy relationships between 
sd-pairs involved in transitions.
As detailed in Fig.~\ref{fi:algo:phase2},
the algorithm linearly sweeps on time instants corresponding to transition 
endpoints and, for every instant $t$ and every extended address $\cal 
A$ (lines~\ref{alg:endpoints} and~\ref{alg:extended-address}),
updates a set $S_{\cal A}$ of sd-pairs $i$ corresponding to active 
transitions that are empathic with each other because they have $\cal A$ in their 
changed set $\Delta(\tau_i)$ (line~\ref{alg:empathic-pairs}). Sets $S_{\cal A}$, 
as well as the time instants $t_{\cal A}$ at which they are updated, are kept 
in special variables which allow access to the last 2 assigned values using 
operators $prev()$ and $pprev()$.
When the size of each $S_{\cal A}$ reaches a local maximum at time $t$ 
(line~\ref{alg:local-maximum}), the algorithm reports a 
candidate event. This corresponds to seeking for the time instant 
at which the highest number of sd-pairs have seen IP address $\cal A$ 
(dis)appear in their traceroute paths. The interval $[prev(t_{\cal A}), 
t_{\cal A}]$ of validity of the local maximum 
(line~\ref{alg:add-candidate-event}) 
is a good candidate for being the time window within which an event has occurred.
The middle part of Fig.~\ref{fig:algorithm_example} shows a sample output of 
this phase, where each segment represents a candidate event: for each 
extended address appearing in the changed sets of $\tau_a$, $\tau_b$, and 
$\tau_c$, the corresponding sets of 
sd-pairs $S_{1^\textrm{pre}}$, $S_{2^\textrm{post}}$, and $S_{3^\textrm{post}}$ 
that involved that address are constructed and updated. In particular, 
set $S_{1^\textrm{pre}}$ reaches its maximum size at time $t_3$, when extended 
address $1^\textrm{pre}$ is in the changed set of $\tau_a$, $\tau_b$, and 
$\tau_c$, namely IP address $1$ has disappeared for sd-pairs $a$, $b$, and $c$. 
The reported candidate event ends at $t_4$, when the size of $S_{1^\textrm{pre}}$ 
is again reduced: it is therefore $(t_3,t_4,\{a,b,c\},1^\textrm{pre})$. Similar 
considerations apply for the construction of the other two candidate events 
$(t_3,t_5,\{a,b\},2^\textrm{post})$ and $(t_2,t_4,\{b,c\},3^\textrm{post})$.


\emph{Phase 3 -- Event inference}: in this phase, detailed in 
Fig.~\ref{fi:algo:phase3}, candidate events are sieved to build a set of 
inferred events, each consisting of a time window, a scope, a set of involved IP 
addresses (which contains the hub of the event), and a type 
(\textsf{up}/\textsf{down}/\textsf{unknown}).
As a first clean-up step, all candidate events whose set of sd-pairs is 
properly contained in the set of sd-pairs of another candidate event that 
overlaps in time 
are discarded 
(lines~\ref{alg:overlapping-cliques-begin}-\ref{alg:overlapping-cliques-end}). 
In this way, only events with maximal impact are reported.
Afterwards, the algorithm considers groups $\mathit{CEvents}(S,t_1,t_2)$ of 
candidate 
events spanning the same time interval $[t_1,t_2]$ and having $S$ as set of 
sd-pairs (line~\ref{alg:group-cliques}), and 
constructs an inferred event for every set $S$ whose size exceeds a configured 
$\mathit{threshold}$: this filters out events with negligible impact. The 
inferred 
event has the following structure 
(lines~\ref{alg:event-ips}-\ref{alg:event-type-end}): the time interval is 
$[t_1,t_2]$; the scope is $S$; the involved IP addresses are 
the union of the addresses of candidate events in 
$\mathit{CEvents}(S,t_1,t_2)$; and the 
type is inferred based on the labels of the extended addresses of candidate 
events in $\mathit{CEvents}(S,t_1,t_2)$ 
(lines~\ref{alg:event-type-begin}-\ref{alg:event-type-end}).
A sample result of the application of this phase is in the lower part of 
Fig.~\ref{fig:algorithm_example}: candidate events 
$(t_3,t_5,\{a,b\},2^\textrm{post})$ and $(t_2,t_4,\{b,c\},3^\textrm{post})$ 
(segments in the second and third row of phase 2, respectively) are discarded 
because their 
sets of sd-pairs are contained in the one of the overlapping candidate event 
$(t_3,t_4,\{a,b,c\},1^\textrm{pre})$. At this point, there is only one 
set of sd-pairs left, $\{a,b,c\}$: assuming no threshold, the only candidate 
event having such set is reported as an event, which affected IP address $1$ 
(that is also the hub of the event) and whose type is $\textsf{down}$ because of 
the label of $1^\textrm{pre}$.

\begin{figure}
\begin{algorithmic}[1]
\footnotesize
\Require a set $\mathit{CEvents}$ of candidate events produced in phase 2 (see 
Fig.~\ref{fi:algo:phase2})
\Ensure a set $\mathit{Events}$ of tuples $(t_1, t_2, S, \Pi, \mathit{type} )$, 
each 
representing an inferred event occurred between $t_1$ and $t_2$, whose scope is 
$S$, which involved the IP addresses in $\Pi$, and whose type is $\mathit{type}$.

\State $\mathit{Events}$ = $\emptyset$
\For{every pair $e=(t_1,t_2,S,{\cal A})$, 
$\tilde e=(\tilde t_1,\tilde t_2, \tilde S, \tilde{\cal A})$ in 
$\mathit{CEvents}$}\label{alg:overlapping-cliques-begin}
   \If{$e$ and $\tilde e$ overlap in time and $\tilde S \subset S$}
   \label{alg:compare-candidate-events}
      \State remove $\tilde e$ from $\mathit{CEvents}$
      \label{alg:remove-overlapping-event}
   \EndIf
\EndFor\label{alg:overlapping-cliques-end}
\State group candidate events $(t_1,t_2,S,{\cal A})$ in $\mathit{CEvents}$ by 
$S$, $t_1$, and $t_2$
\label{alg:group-cliques}
\For{every computed group $\mathit{CEvents}(S,t_1,t_2)$}
   \label{alg:every-group}
   \If{$|S| > \mathit{threshold}$}
      \State $eaddr$ = $\bigcup_{(t_1,t_2,S,{\cal A})\in 
\mathit{CEvents}(S)}{\cal A}$\label{alg:event-ips}
      \State $\Pi$ = $eaddr$ (without labels)
      \State $type$ = \textsf{unknown}
         \label{alg:event-type-begin}
      \State $type$ = \textsf{down} if all addresses in $eaddr$ are tagged as 
\pre
      \State $type$ = \textsf{up} if all addresses in $eaddr$ are tagged as \post
         \label{alg:event-type-end}
      \State add $(t_1, t_2, S, \Pi, type)$ to $\mathit{Events}$
   \EndIf
\EndFor\label{alg:infer-event-end}
\Return $\mathit{Events}$

\end{algorithmic}
\caption{Phase 3 of our algorithm: it reports inferred events starting from a 
set of candidate events produced in Phase 2.}
\label{fi:algo:phase3}
\end{figure}

Our algorithm is correct and complete, as stated by the 
following theorems.

\begin{theorem}[Correctness]
Each event inferred by our algorithm corresponds to a physical event.
\end{theorem}

\begin{IEEEproof}
Let $(t_1,t_2,S,\Pi,type)$ be an inferred event.
By construction, every address 
$\pi$ in $\Pi$ has (dis)appeared in all transitions $\tau_i$ for every $i\in 
S$, and $S$ has maximal size: therefore $\pi$ is a candidate for being the hub 
of a physical event.
Moreover, since $\pi$ (dis)appears in traceroutes in the interval in which 
transitions $\tau_i$ intersect, namely between $t_1$ and $t_2$, this is also the 
time window in which the event has occurred.
\end{IEEEproof}

\begin{theorem}[Completeness]\label{the:completeness}
For every visible physical event, an inferred event is reported by our 
algorithm.
\end{theorem}

\begin{IEEEproof}
Suppose a physical event $E^\downarrow$ with hub $h$ occurs at time $\bar 
t$: the traceroutes for all sd-pairs $i$ that are in the scope $S(E^\downarrow)$ 
of $E^\downarrow$ will therefore change after $\bar t$, and phase 1 of the 
algorithm constructs transitions $\tau_i$ whose intervals contain $\bar t$. All 
such transitions must intersect at a common interval $[t_1,t_2]$ comprising $\bar 
t$ and have $h^\textrm{pre}\in\Delta(\tau_i)$. By the definition of scope, the 
cardinality of set $S_{h^\textrm{pre}}$ reaches a local maximum between $t_1$ 
and $t_2$ in phase 2 of the algorithm, and a candidate event 
$e=(t_1,t_2,S(E^\downarrow),h^\textrm{pre})$ is thus constructed. Set 
$S(E^\downarrow)$ is the largest possible set of sd-pairs affected by 
$E^\downarrow$, therefore $e$ is not filtered in phase 3 and an event 
$(t_1,t_2,S(E^\downarrow),\Pi,type)$ with $h\in\Pi$ is reported.
Analogous arguments can be applied to the case of an event $E^\uparrow$.
\end{IEEEproof}

The computational complexity of our inference algorithm is $O(|T| +  
|\mathit{CEevents}|^2\cdot I)$, where $T$ is the set of transitions and $I$ is 
the maximum impact. In fact, 
phases 1 takes $O(|T|)$. Since the size of the changed set 
of every transition  is bounded by the maximum length of traceroute paths and 
sets $T_{\cal A}$ and $sdp$ can be updated during the sweep, phase 2 also 
takes $O(|T|)$. Phase 3 
takes $O(|\mathit{CEvents}|^2\cdot I)$ because of the overlap check at 
lines~\ref{alg:overlapping-cliques-begin}-\ref{alg:overlapping-cliques-end} (the 
following event construction can be performed efficiently by scanning candidate 
events).

Several issues are inherent in using real-world traceroute data, but our 
model and algorithm can effectively cope with them, as also confirmed by 
experimental results.
First of all, a single network device equipped with multiple network interfaces 
(e.g., a router) may reply with different IP addresses in different traceroutes, 
a phenomenon known as \emph{aliasing}~\cite{marchetta2013pythia}. As a 
consequence, detection of some empathies may fail, causing 
our algorithm to infer multiple small events rather than a single larger one in 
the worst case. Delays in the propagation of routing changes (for example due to 
routing protocol timers) have a similar effect on the algorithm's output, which 
may include multiple copies of the same event with slightly different time 
intervals and scopes.
On the other hand, multiple simultaneous events can interfere 
with each other, namely the corresponding transitions may overlap and their 
changed sets may have elements in common. 
Such events can still be detected by our algorithm, even if their scope can only 
be identified with a limited precision. Under rare circumstances, some fictitious 
or improperly time-skewed events may also be inferred.
In practice, none of these cases prevents our algorithm from reporting
events, and their incidence was negligible in our\nolinebreak{} experiments.

One aspect that may indeed taint the output of our algorithm is that the vast 
majority of Internet paths traverse load balancers~\cite{augustin2007paris}: 
they are the cause of a high number of apparent routing changes which may be 
improperly reported as physical events.
Compensating this issue requires knowledge of the load balancers, which is 
realistic for an Internet Service Provider that wants to apply our methodology, 
and 
can otherwise be constructed by applying discovery techniques such as Paris 
Traceroute~\cite{augustin2006avoiding}. Unfortunately, this technique was not 
yet available in the measurement networks we considered, therefore we 
preprocessed traceroutes by using a simple heuristic that cleaned up most 
of the noise introduced by load balancers: we analyzed all the input traceroutes 
in their time order and, for each destination, we tracked the evolution over time 
of the routing (actually the next hop) of every node along the traceroute paths. 
Nodes with unstable routing (i.e., that change the next hop in more than 
20\% of the samples) are considered to belong to a load balancer and their next 
hops are replaced by a single arbitrarily chosen representative IP address.
\section{Experimental Results}\label{sec:experiments}

We executed our algorithm on several sets of traceroute paths collected by 
currently active measurement networks, with the intent to verify that the 
inferred events matched physical events.
We first considered traceroutes affected by a sequence 
of routing changes injected with a known schedule, used as ground 
truth. We then used our algorithm to detect spontaneous events happened in the 
network of a European operator.

\subsubsection{Induced Event Analysis}.

For this experiment we partnered with an Italian ISP that has BGP peerings with 
three main upstream providers and with a number of ASes at three Internet 
eXchange Points (IXPs), i.e. MIX, NaMeX (the main IXPs in Italy), and 
AMS-IX\footnote{NaMeX and AMS-IX are connected by a link. However, it was not 
used by any means in our specific setting.}.
An IP subnet reserved for the experiment was announced via BGP to different 
subsets of peers, according to the schedule in 
Table~\ref{tab:seeweb-experiment-results}.
During the experiment, 89 RIPE Atlas probes located in Italy were instructed to 
perform traceroutes every 10 minutes (between 2014-05-02 13:00 UTC and 2014-05-03 
15:00 UTC)  targeting a host inside the reserved subnet. After applying the 
load balancers cleanup heuristic described in Section~\ref{sec:methodology}, we 
fed our algorithm with the collected traceroute measurements.

\begin{figure}
   \centering
   \includegraphics[width=.6\columnwidth]{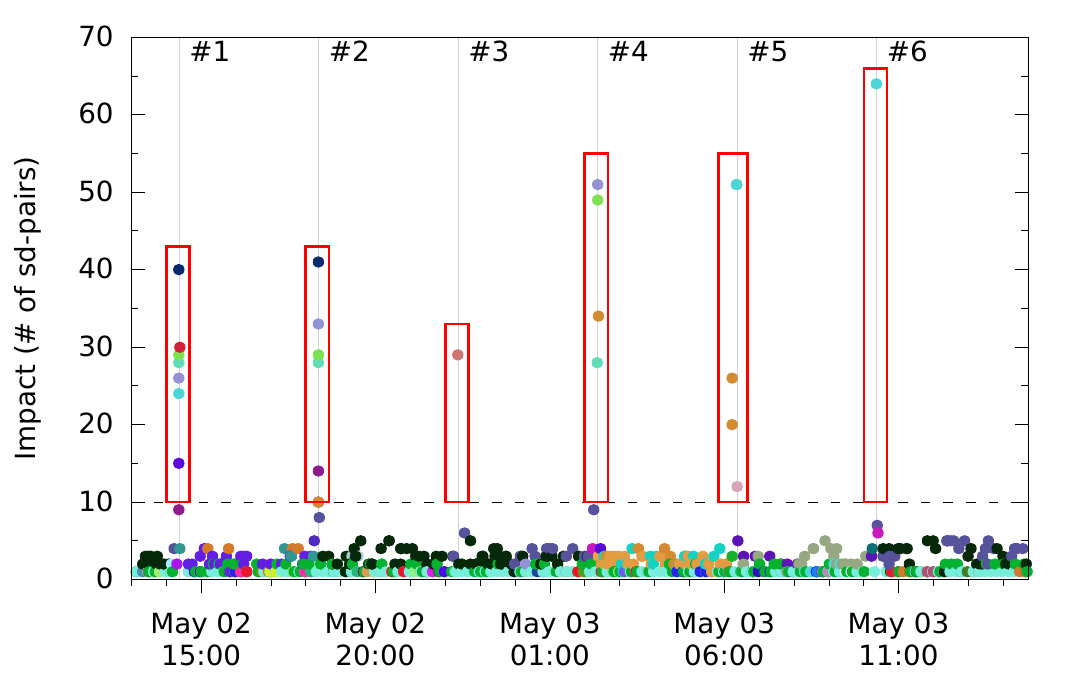}
   \caption{Impacts of the events inferred during experiment 1 (induced 
events).}
   \label{fig:events-impact-experiment1}
\end{figure}

The produced output, which took only a few seconds to compute, is plotted in 
Fig.~\ref{fig:events-impact-experiment1}: 
each inferred event $(t_1,t_2,S,\Pi,\mathit{type})$ is represented by a point 
whose coordinates are the center of interval $[t_1,t_2]$ (X axis) and the event's
impact $|S|$ (Y axis), and whose color identifies a specific set $\Pi$ of 
involved IP addresses. 
Out of all the inferred events, $23$ exceeded the impact threshold of $10$ 
(dashed horizontal line in the figure), which clearly separates them from 
background noise.
It is evident that these $23$ events tend to concentrate (red boxes) around 
the time instants of BGP announcements (vertical gray lines numbered according 
to the rows of 
Table~\ref{tab:seeweb-experiment-results}), and indeed the 
center of the time interval of each event falls within seconds from the 
corresponding announcement. In addition, the maximum extension of each interval 
$[t_1,t_2]$ was $2$ minutes, confirming that our 
methodology can detect an event very quickly after the instant in which it 
\mbox{actually happened}.
Set $\Pi$ consisted of a single IP address for $87\%$ of the events and of at 
most $4$ IP addresses for $2$ events, demonstrating a high precision in pointing 
out possible event causes.

For at least one announcement change (\#3) the detection was optimal, namely we 
inferred a single event where all the $29$ involved sd-pairs switched from MIX 
to NaMeX. Multiple events were instead inferred in the other cases, due to 
asterisks in traceroutes or interference between routing changes happening 
close in time to each other. One of the inferred events even allowed us to 
discover an undeclared backup peering whose existence was later confirmed by  
the ISP.

\begin{table}
\centering
\caption{Schedule of BGP announcements for the controlled experiment in 
Section~\ref{sec:experiments}.}
\begin{tabular}{|c|c|c|c|c|c|}

\cline{2-6}
\multicolumn{1}{c|}{} & \textbf{Time} & \textbf{Upstreams} & 
\textbf{MIX} & \textbf{NaMeX} & \textbf{AMS-IX} \\
\cline{2-6}
\multicolumn{1}{c|}{} & May 02, before 14:22 & \checkmark & \checkmark & 
\checkmark & \checkmark \\
\hline
\#1 & May 02, 14:22 & \checkmark & & & \\
\hline
\#2 & May 02, 18:22 & & \checkmark & \checkmark & \\
\hline
\#3 & May 02, 22:22 & & & \checkmark & \\
\hline
\#4 & May 03, 02:22 & & \checkmark & & \\
\hline
\#5 & May 03, 06:22 & & & & \checkmark \\
\hline
\#6 & May 03, 10:22 & \checkmark & \checkmark & \checkmark & \checkmark\\
\hline

\end{tabular}
\label{tab:seeweb-experiment-results}
\end{table}

\begin{figure}
   \centering
   \includegraphics[width=.6\columnwidth]{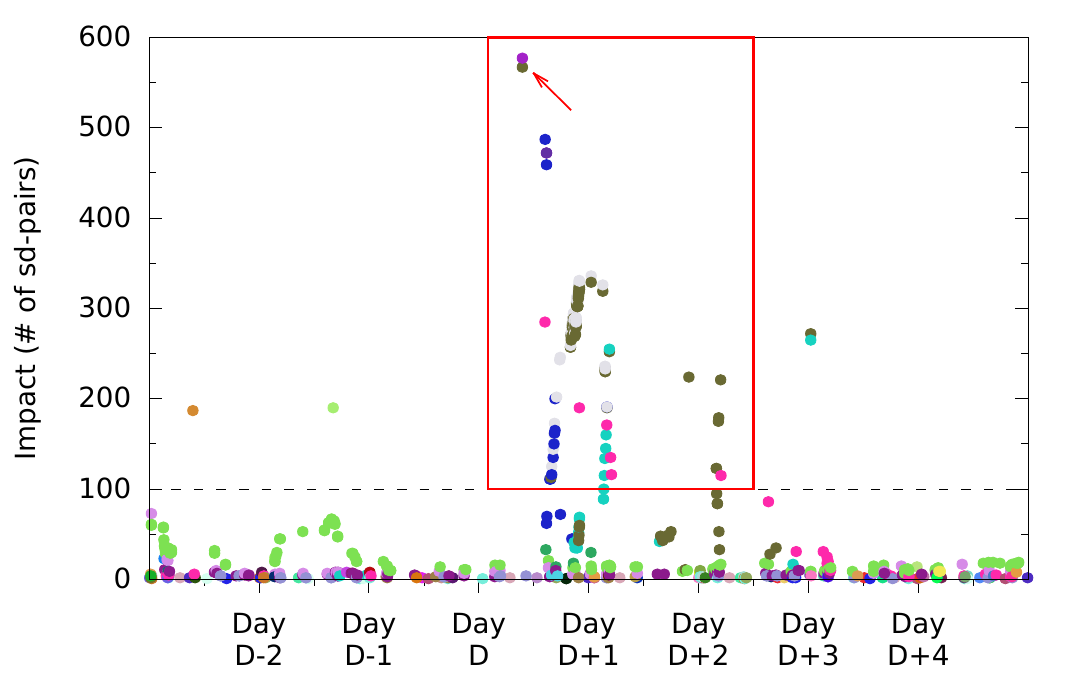}
   \caption{Impacts of the events inferred in experiment 2 (spontaneous 
events).}
   \label{fig:events-impact-experiment2}
\end{figure}

\subsubsection{Spontaneous Event Analysis}

For the second experiment, we considered traceroute paths collected every $8$ 
hours by $3320$ probes distributed within the network of a European operator, 
called \eo in 
the following for privacy reasons. Traceroutes were performed towards 
destinations located both inside and outside \eo's network.
In a private communication 
\eo informed us about a ``routing failure'' in one of its ASes occurred on day 
$D$, therefore we focused on traceroutes collected in a $9$-days 
time window comprising this day.
We applied our load balancers cleanup heuristic on a slightly richer data set 
consisting of almost $260.000$ traceroutes collected over $15$ days. Our 
algorithm then took about $3$ minutes to completely process the cleaned set of 
traceroutes, computing almost $60.000$ transitions in phase 1. We separately 
ascertained that the load balancers heuristic reduced the number of 
transitions by almost a factor of $3$ and the number of inferred events by almost 
a factor of $20$.
For an improved accuracy, we filtered out inferred events that did not 
involve (in $\Pi$) any IP addresses within \eo's network, obtaining the events 
in Fig.~\ref{fig:events-impact-experiment2}. 

Considering the average impact of the inferred events, we 
set the threshold at $100$ (dashed horizontal line in the figure).
The figure shows that events exceeding this threshold are mainly concentrated 
within a time window whose center falls within 24 hours from Day $D$, 
and are followed by some less impactful events occurring up to 2 days later 
(red box in the figure), totaling $199$ events involving $838$ unique sd-pairs. 
Our algorithm also singled out their candidate causes pretty accurately, given 
that the union of all sets $\Pi$ consisted of as few as $7$ IP addresses.
Considering the frequency of traceroutes, these events were also somewhat 
precisely located in time: the length of their time intervals ranged from about 
10 hours to as low as 1 second (due to traceroutes not being synchronized), with 
a standard deviation of 30 minutes.

As it can be seen from the figure, reported events are rather fragmented 
despite affecting the same set of IP addresses (points with 
the same colors 
in the figure): this is due to the fact that routing propagation delays caused 
many non-overlapping transitions to be constructed in phase 1. Interestingly, 
the two events with impact higher than $550$ (indicated by an arrow in the 
figure) were of type \textsf{down} and \textsf{up}, indicating that all the 
traceroute paths of the involved sd-pairs switched to alternate routes sharing 
some common IP addresses (all within \eo's network).
Events whose time window is centered between $D+1$ and $D+2$ are likely due to 
configuration changes undertaken to restore a working routing.

After submitting our results to \eo, they confirmed that inferred events with 
outstanding impacts had very good match with the incident, 
and subsequent events corresponded to actions aimed at restoring the full 
operational state.

We believe these experiments highlight the effectiveness of our methodology in 
detecting events, regardless of whether they are induced and recurring 
(experiment 1) or spontaneous and isolated (experiment 2), and there is 
evidence that detection can happen very quickly after the occurrence of an 
event.

\section{Conclusions and Future Work}\label{sec:futurework}

We have presented a model and methodology for the identification and analysis of 
network events based on the notion of empathic traceroute measurements. We have 
translated our theoretical approach into an algorithm and applied it to 
real-world data, proving the effectiveness of our methodology.

We plan to further validate our approach with other measurement platforms (see 
Section~\ref{sec:related-work} for examples), topologies, and network events. We 
will focus in particular on intra-domain routing events as opposed to BGP 
routing changes.
Further, we will study heuristics to merge two or more inferred events that are 
likely to represent one single network event, and work on devising an on-line 
version of our algorithm, which could effectively integrate the core of an 
alerting system.

\bibliographystyle{IEEEtran}
\bibliography{bibliography}

\end{document}